\documentclass[12pt]{iopart}
\usepackage[dvips]{graphicx}
\begin{document}
\title{Magnetic field induced metastability in the self-doped manganites}
\author{K. De, S. Majumdar, S. Giri$^*$}

\address{Department of Solid State Physics, Indian Association for the Cultivation of Science, Jadavpur, Kolkata 700 032, India}
\ead{$^*$sspsg2@iacs.res.in}
\begin{abstract}
We observe an interesting scenario of field induced metastabilities in the resistivity measurements. The resistivity exhibits a bifurcation of zero-field cooled and field-cooled (FC) temperature dependence below the paramagnetic to ferromagnetic transition ($T_c$). The clear evidence of {\it inverse} thermal hysteresis of resistivity is observed under FC condition in between 85 and 200 K, indicating the first order phase transition at $T_c$. The additional observation of minor hysteresis loops confirms the phase coexistence. The interesting features in the dynamics of resistivity is  observed at 40 K, which is suggested due to the effect of domain wall dynamics.  
\end{abstract}

\maketitle

The mixed valent manganites with perovskite structure are fascinating  due to the wealth of  physical properties  arising from  interplay between  structural, electronicn, and magnetic properties of the system \cite{salamon,dagotta}. The  co-existence of different competing factors often gives rise to  metastable behaviour, which are evident from  temperature  rate dependence \cite{levy1}, unusual relaxation dynamics \cite{wu}, frequency dependence \cite{kde1}, relaxor ferroelectric-like behaviour \cite{dho}, memory effects in various physical properties of manganite compounds \cite{levy1,quin}. The existence of phase separation  is commonly noticed in these systems, where  submicrometer ferromagnetic (FM) metallic and non-ferromagnetic insulating state can coexist. The insulating phase  can be either a charged ordered (CO)/orbitally ordered AFM sate or a paramagnetic state. The phase separation (PS) scenario associated with the out of equilibrium state has been reported in the low hole doped and CO manganites. Recently, the  magnetotransport properties with the PS scenario has also been reported in the self doped compound, La$_{1- \delta}$MnO$_x$, where the interplay among electronic, magnetic, orbital, and structural properties are strongly dependent on $\delta$ and {\it x} \cite{troyanchuk2, sankar}. In case of La deficiency, FM orbitally disordered phase appears in presence of AFM orbitally ordered phase associated with the structural instabilities in La$_{1- \delta}$MnO$_x$ depending on the values of $x$ \cite{troyanchuk2}. The dynamics of zero-field cooled (ZFC) magnetization as well as in the ac susceptibility have recently been reported indicating the metamagnetic states, which were suggested by the effects of domain wall dynamics in La$_{1- \delta}$MnO$_3$ \cite{sankar, muroi}. The extensive studies in the self doped manganites have not been performed as compared to  hole doped and CO manganites and it demands special attention because of the intriguing transport and magnetic properties. 

\par
In this letter, we report mainly on the resistivity ($\rho$) measurements on La$_{0.87}$Mn$_{0.98}$Fe$_{0.02}$O$_x$ under different conditions of magnetic field and temperature cycling, demonstrating interesting features of field induced metastabilities viz., field cooled (FC) effect, thermal hysteresis, and interesting dynamical features. In addition, we incorporate a simple experimental technique of minor hysteresis loops to identify the phase coexistence in the self doped manganites. The evidence of metastable states in different manganites have been reported quite extensively based on the magnetization studies. However, the examples of such metastable feature in the resistivity measurement are rare, which is reported in this article for self doped manganites. 

\par
The polycrystalline sample with nominal composition, La$_{0.87}$Mn$_{0.98}$Fe$_{0.02}$O$_x$ was prepared by chemical route \cite{kde2}. The final heat treatment was performed at 1200 $^0$C for 12 h in air, followed by furnace cooling down to room temperature. The magnetoresistance (MR) was measured by  standard four-probe technique, fitted with an electromagnet. The magnetization ($M$) was measured using a commercial superconducting quantum interference device (SQUID) magnetometer. In ZFC measurements of $\rho$ and $M$, the sample was cooled down to the desired temperature at zero  magnetic field and the measurements were performed in the heating cycle with the applied  magnetic field. For the FC condition,  the sample was cooled down to the desired temperature with magnetic field and the measurements were performed while heating.  The rate of temperature in all the thermal variation of $\rho$ was performed  at a  rate of 2 K/min.

\begin{figure}
\centering
\includegraphics[width = 9 cm]{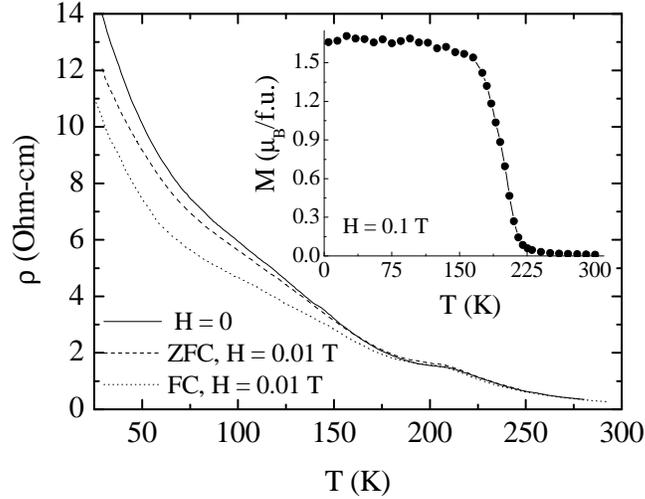}
\caption{Temperature dependence of resistivity under zero magnetic field, zero-field cooled and field-cooled conditions. The inset shows the temperature dependence of magnetization in zero-field cooled condition under 1.0 kOe}
\end{figure}

	The powder XRD analysis confirms the single phase of the compound, which could be indexed by the monoclinic structure ({\it I}2/{\it a}). Temperature dependence of $M$ in ZFC condition under 1.0 kOe is shown in the inset of figure 1, indicating a paramagnetic to ferromagnetic transition ($T_c$) around $\sim$ 205 K. The monoclinic distortion of the crystal structure and high value of $T_c$ suggest higher oxygen stoichiometry ($x \geq$ 2.9) in accordance with the reported results \cite{troyanchuk2,kopcewicz}. Temperature dependence of $\rho$ in zero field, ZFC, and FC conditions are shown in figure 1. In absence of field, an anomaly in $\rho$ is observed around $\sim$ 209 K close to $T_c$. The semiconducting like $T$ dependence is observed below $T_c$ with a change of slope around $\sim$ 75 K. The MR is noticed below $\sim$ 200 K, which increases with decreasing temperature. A striking observation of bifurcation of $\rho$ in the temperature dependence under ZFC and FC conditions is clearly observed below $\sim$ 200 K. The bifurcation of ZFC and FC data  has commonly been found in the temperature dependence of magnetization, when the  metastability arises from spin freezing \cite{mydosh}. The evidence of FC effect in $\rho$ is rather, a rare example among transition metal oxides, which has recently been reported even in few manganites \cite{dho,smolyaninova}. A strong FC effect in $\rho$ was observed in the phase separated (La$_{0.8}$Gd$_{0.2}$)$_{1.4}$Sr$_{1.6}$Mn$_2$O$_7$ below spin glass (SG) temperature ($T_{SG}$), where glassy behaviour was ascribed to the competing magnetic interactions between FM and AFM states \cite{dho}. The other kind of example of ZFC-FC effect in $\rho$ was observed in CO compounds, where the CO state in the system is destroyed by the application of strong magnetic field, resulting in the different $T$ dependence of FC and ZFC resistivities below the charge ordering temperature ($T_{CO}$) \cite{smolyaninova}. In the present case, the origin of FC effect does not meet any of the above criteria rather, it exhibits another example of FC effect in $\rho$ in self doped manganites. 

\par	
To understand further about the FC effect, resistivity and magnetization were measured by varying the magnetic field at selected temperatures under ZFC and FC conditions as seen in figure 2. The non-linearity of the low-field MR curves at 150 and 100 were noticed below ~ 0.6 and ~ 1.0 kOe, respectively, while the low-field MR curve at 170 K is different than others as seen in figure 2 (top panel). In accordance with the $T$ dependent results a small difference between ZFC and FC curves were noticed at 40, 100, 150, and 170 K in the region of almost linear field dependence. The values of MR, defined as [$\rho$(0) - $\rho$($H$)]/$\rho$(0), are ~ 10, ~ 12, ~ 15, and ~ 20 \% at 170, 150, 100, and 40 K, respectively under 5.0 kOe. Low-field MR under both ZFC and FC conditions and magnetization curves under ZFC condition are shown in figure 2 (bottom panel), where non-linearity is noticed till ~ 1.5 kOe for both the cases. The inset shows the magnetization curve up to 20 kOe, exhibiting the typical features of soft ferromagnet.

\begin{figure}
\centering
\includegraphics[width = 9 cm]{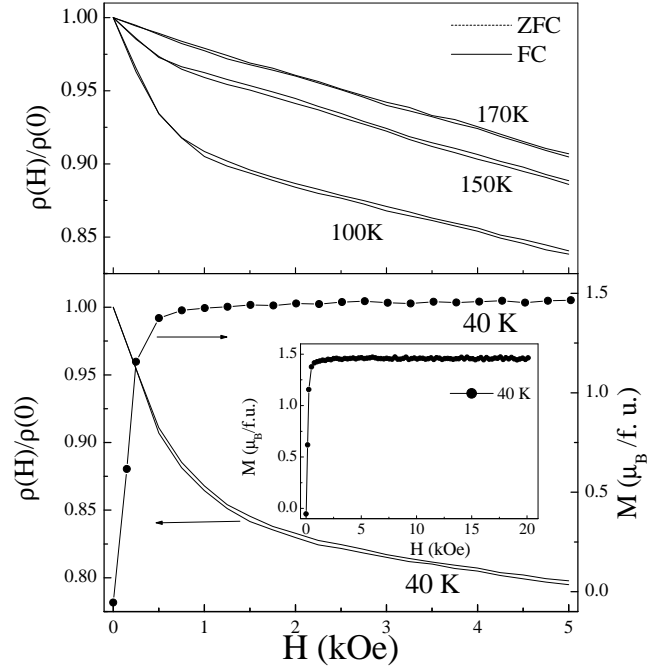}
\caption{The top panel shows the field dependence of resistivity, $\rho$($H$) scaled by the resistivity under zero field, $\rho$(0) under zero-field cooled (ZFC) and field-cooled (FC) conditions at 170, 150 and 100. The bottom panel depicts field dependence of resistivity under ZFC and FC conditions and magnetization under ZFC condition at 40 K in the low field region. The inset of (b) shows the magnetization curve at 40 K up to 20 kOe.}
\end{figure}

\par
The striking signature  of metastability was observed by the thermal hysteresis in $\rho$, when the measurement was performed in presence of magnetic field under the following scheme of $T$ variation. The sample was first cooled down to 25 K, $\rho$ was then measured in the heating cycle up to 275 K ($>>$ $T_c$), and finally, $\rho$ was remeasured in the cooling cycle down to 25 K. The above scheme of measurements was performed both in absence and presence magnetic field. The thermal hysteresis in $\rho$ was not observed in zero magnetic field as seen in the inset of figure 3(a). On the other hand, a clear evidence of thermal irreversibility in $\rho$ is noticed below $T_c$ [figure 3(a)], when the measurement was performed in FC condition under 1.0 kOe.  Interestingly, the observed thermal hysteresis is {\it inverted} in nature, i.e, the heating leg lies below the cooling leg, despite the fact that $\rho(T)$ increases with decreasing temperature. In case of a normal  case of thermal hysteresis,  the physical parameter under investigation  tends to retain the high temperature or low temperature value, when the temperature is decreased or increased  respectively. The bifurcation between cooling and heating curves occurs around $T_H$ = 200 K, which is close to $T_C$,  and they join up  around $T_L$ = 85 K. We performed several successive cycles of heating and cooling in the FC condition with 1.0 kOe of applied field and the observed inverted hysteresis loops were reproducible. Since, the measurements are performed using an electromagnet, the question of experimental artifact due to persistent current in the superconducting magnet can be ruled out. To further confirm the existence of this hysteresis, we carefully measured minor hysteresis loops (MHL) in the region of hysteresis both on the heating and cooling legs \cite{smajumdar}. The MHL were obtained by suddenly reversing the direction of temperature at an intermediate temperatures within the  region of hysteresis and then the measurements were performed up to $T_H$ or $T_L$. The examples of MHL in the cooling cycles and MHL in the heating cycles are shown in Figs. 3(b) and (c), respectively. The clear evidence of MHL not only confirms the hysteresis, but also support phase coexistence, atleast, across the region of hysteresis. The existence of thermal hysteresis in $M$ as well as in $\rho$ were observed for charge order  manganites, indicating the first order transition below $T_{CO}$, which have been observed even in zero magnetic field \cite{smolyaninova}. In the present observation, the thermal hysteresis is also ascribed to the field induced first order phase transition associated with the PM to FM transformation at $T_c$. In fact, the structurally driven first order phase transition along with the PM to FM transitions were reported for the monoclinic phase of La$_{0.88}$MnO$_x$ (2.92 $ \leq x \leq 2.96$), though the first order phase transition was predicted based on the qualitatively from the analysis of magnetization curves around the transition temperature \cite{troyanchuk2}. 

\begin{figure}
\centering
\includegraphics[width = 9 cm]{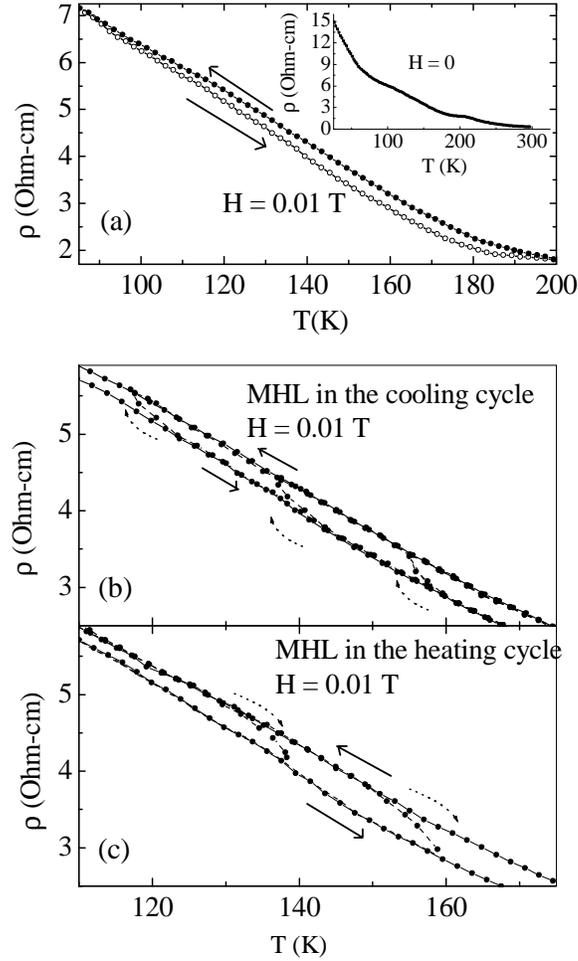}
\caption{ Thermal hysteresis of resistivity under field-cooled condition (a), minor hysteresis loops in the cooling cycles (b), and in the heating cycles (c).  The inset of (a) exhibits no thermal hysteresis in zero field.
Arrows indicate the mode of thermal cycling.}
\end{figure}

To understand  insight into the nature of metastability, we measured the dynamics of $\rho$ in the measured temperature range. We did not observe any noticeable change in $\rho$ with time ($t$) in case of isothermal holding in zero magnetic field for any temperature down to 10 K. Even in presence of field, time dependence of $\rho$ is absent in the region of thermal hysteresis (200-85 K). However, we noticed the change in $\rho$ with $t$ convincingly at 40 K, which is well below the thermal irreversibility and the change of slope of resistivity at $\sim$ 75 K (figure 1). In order to  shed more light on the time dependent character, we investigated the aging effect with the following experimental protocol. The sample was first cooled down to 40 K from 270 K  under ZFC condition, waited for different time period ($t_w$), and then, $\rho$ was measured with $t$ just after switching on the field. The dynamics in $\rho$ was measured under 1.0 and 2.0 kOe, where the MR curve exhibits nonlinear and almost linear dependence, respectively (figure 2). The $t$ dependence associated with different characteristic features in $\rho$ were observed at different fields as shown in figure 4. Note that a sharp change in $\rho$ is surprisingly noticed after a time period roughly equal to $t_w$ = 60 and 120 minutes under 1.0 kOe, while the anomaly in $\rho$ around $t_w$ is absent under 2.0 kOe. Nevertheless, the common feature in the dynamics of $\rho$ under 1.0 and 2.0 kOe show that $\rho$ vs $t$ plots do not follow the same path at different $t_w$, like a aging effect in magnetization for the SG compounds. 

\par
The aging reflects the cooperative relaxation process. In case of canonical SG compound, the aging effect is commonly observed below $T_{SG}$, where the relaxation rate of magnetization, $S$ = 1/$H$($dM$/$d$ln$t$), is found to be maximum around, $t$ = $t_w$ \cite{mydosh}. The evidence of aging in $\rho$ is rather, a rare example even in mixed valent manganites. Recently, the evidence of sharp change in $\rho$ at $t_w$ was also reported by Wu {\it et al}. \cite{wu}, suggesting the glassy transport phenomena in the phase separated cobalite. In accordance with this results, the sharp change in $\rho$ is observed directly in the $\rho$ {\it vs.} $t$ plot, not in the relaxation rate of $\rho$. Note that the $\rho$ in manganites are sensitive to the polarization of both the metastable state and FM clusters. The metastable states may appear either from SG state or domain wall dynamics, which can be distinguished from the low field measurements. On the other hand, $\rho$ is mainly dominated by the FM clusters in case of high field measurements. The sharp change of $\rho$ at $t_w$ under 1.0 kOe indicates the dominant character of metastable component, while such an anomaly is smeared out at 2.0 kOe, because of the dominant influence of polarization of the FM clusters. The above feature is rather consistent with the magnetization and MR curves as seen in figure 2 (bottom panel), where the saturation of magnetization as well as the linearity of MR are noticed above 1.5 kOe. The phase co-existence of orbitally ordered AFM and orbitally disordered FM states were already reported for self doped manganites \cite{troyanchuk2}. In case of minor substitution of Fe, Fe$^{3+}$ occupy Mn$^{3+}$ site randomly, which may introduce the frustration in the system. Thus, the frustration arising from different competing factors can give rise to the glassy behaviour, which may be responsible for the sharp change in $\rho$ at $t_w$. The other possibility of the origin of sharp change in $\rho$ at $t_w$ is electron scattering correlated to the the domain wall dynamics, which is rather more plausible than glassy behaviour. Note that the magnetization curve at 40 K exhibits the feature of typical soft ferromagnet, where the magnetization saturates at a very low magnetic field of 1.5 kOe. In the present observation, the system was cooled down to 40 K, waited for $t_w$ and then magnetic field was suddenly switched on just before the start of measurement. In such a case the system keeps the tracks of domain wall movement in to the memory till $t_w$ during waiting at 40 K, which is reflected in the relaxation measurement, where the rate of MR discontinuously changes at $t_w$. The domain walls have a significant influence on the electronic conduction in FM metals \cite{dugaev}. Recently, the influence of domain wall on the resistivity has been found to contribute  significantly  in different FM materials {\it viz.}, epitaxial thin film \cite{kent}, (Ga,Mn)As \cite{chiba}, nanowires of Ni80Fe20 and Ni \cite{lepadatu}. To the best of our knowledge, the influence of domain wall dynamics on $\rho$ is not reported so far for ferromagnetic manganite. However, the slow relaxation dynamics in the magnetization and ac susceptibility have recently been reported for La$_{1-x}$MnO$_3$ by several authors, which were suggested due to the domain wall dynamics \cite{sankar, muroi, muroi2}.

\begin{figure}
\centering
\includegraphics[width = 9 cm]{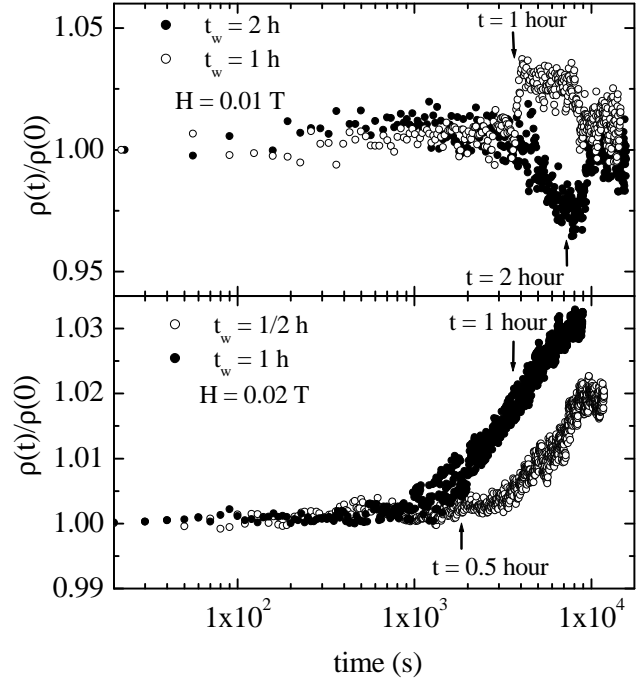}
\caption{ Time dependence of resistivity measured under zero-field cooled condition of 1.0 kOe field, indicating the sharp change of resistivity around different $t_w$ (a) and under 2.0 kOe without any anomaly at $t_w$. Arrows indicate the time at $t_w$.}
\end{figure}

\par
The observation of thermal hysteresis in $\rho$ confirms the first-order magneto-structural phase transition at $T_c$. It is difficult to confirm the origin of such field induced first order transition from this macroscopic transport measurements. However, the transition might be a weakly first order in absence of applied magnetic field, but in presence of field it shows first-order nature. Note that the existence of local clustering and structural disorder were pointed out by Troyanchuk {\it et al}. to establish an intrinsic chemical and structural inhomogeneity on the nanometer scale in the orthorhombically distorted and orbitally disordered insulating FM state in La$_{0.88}$MnO$_x$ (2.86 $ \leq x \leq 2.91$) \cite{troyanchuk2}. Similar kind of scenario of structural inhomogeneity is suggested in case of minor substitution of Fe, though the monoclinic phase is noticed from the powder X-ray diffraction. In fact, recent theoretical model \cite{dagotta} supported by the experimental observation in low hole doped La(Sr)MnO$_3$ \cite{shibata} also exhibits the existence of intrinsic structural inhomogeneity in manganites. Here, the value of $\rho$ under FC condition reduces from the zero field equilibrium value below $T_c$, which allows us to suggest that the size of the FM clusters consisting of orbitally disordered state increase under magnetic field, in consistent with the other phase separated manganites \cite{levy1,wu,dho}. The growth of FM cluster might be related to the structural transformation below $T_c$, in addition to the orbital dynamics in the thermal irreversible region \cite{kopcewicz}.
The observed {\it inverse} hysteresis [figure 3(a)] can be ascribed to the percolative conduction \cite{stauffer} through the orbitally disordered FM metallic and orbitally ordered AFM insulating phases, very similar to CO compounds \cite{salamon,levy1,wu,dho,smolyaninova}. Although, the metallic ($d\rho/dT > 0$) FM clusters start to grow below $T_C$, the overall resistivity remains semiconducting ($d\rho/dT < 0$) in nature down to the lowest temperature of measurement due to the dominant character of insulating phase. Thus, the total resistivity, $\rho_{tot}(T)$ will have two components, FM metallic part, $\rho_M(T)$ and insulating part, $\rho_I(T)$, with $\rho_{tot}(T) = \rho_M(T) \oplus \rho_I(T)$. The metallic component decreases with {\it T}, which shows lower value in the heating cycle than that of the value in the cooling cycle in case of thermal hysteresis. If we believe that the metallic component of resistivity only shows the hysteresis, it will be reflected as inverse hysteresis on the resulting semiconducting $\rho_{tot}(T)$ behaviour. 

In summary, we have presented the evidence of metastabilities in the resistivity under FC condition, while such metastabilities do not exist in absence of magnetic field. The thermal hysteresis in resistivity are observed below $T_c$ under FC condition, which suggests the first order phase transition at $T_c$. The additional observation of MHL suggests the phase coexistence. The sharp changes in $\rho$ associated with the characteristic features of relaxation dynamics at 40 K are suggested due to the domain wall dynamics. 
	
One of the authors (S.G.) wishes to thank DAE, India for the financial support. The magnetization data using SQUID magnetometer were measured under the scheme of Nanoscience and Nanotechnology initiative of DST at IACS, Kolkata, India.

\end{document}